\documentclass[twocolumn]{aa}

  \usepackage{amscd}
  \usepackage{amsmath}
  \usepackage{amssymb}
  \usepackage[T1]{fontenc}
  \usepackage[varg]{txfonts}
  \usepackage{graphicx}
  \usepackage{natbib}
  \usepackage{verbatim}
  \usepackage{array}

  \newcommand{\mbf}[1]{\ensuremath\mbox{\boldmath{$#1$}}}

\begin{document}

\title{Full-star Type Ia supernova explosion models}

\author{F. K. R{\"o}pke\inst{1},
        \and
        W. Hillebrandt\inst{2}}     
   \offprints{F. K. R{\"o}pke}

   \institute{Max-Planck-Institut f\"ur Astrophysik,
              Karl-Schwarzschild-Str. 1, D-85741 Garching, Germany\\
              \email{fritz@mpa-garching.mpg.de}
              \and
              \email{wfh@mpa-garching.mpg.de}
             }

\abstract{We present full-star simulations of Type Ia
  supernova explosions on the basis of the standard Chandrasekhar-mass
  deflagration model. Most simulations so far considered only one spatial
  octant and assumed mirror symmetry to the other octants. Two
  full-star models are evolved to homologous expansion and compared with
  previous single-octant simulations. Therefrom we analyze the effect of
  abolishing   the artificial symmetry constraint on the evolution of
  the flame  surface. It turns out that the development of
  asymmetries depends on the chosen initial flame configuration.
  Such asymmetries of the explosion process could possibly contribute
  to the observed polarization of some Type Ia supernova spectra.
}

\maketitle


\section{Introduction}
\label{intro_sect}

Recent observations have shown that spectra of---at least some---Type
Ia supernovae (SNe Ia) show significant polarization
\citep[e.g.][]{howell2001a,wang2003a}. The origin of this effect is still unclear,
but some possible
mechanisms have been proposed
\citep{wang2003a,kasen2003a,kasen2004a}. Besides peculiar 
spatial structures of the remnant and interaction of the ejecta with
the companion star, anisotropies of the explosion process itself could
contribute to the observed polarization.

This idea fits well into the currently favored astrophysical model of
SNe Ia. These events are assigned to thermonuclear
explosions of carbon/oxygen white dwarf (WD) stars in binary
systems \citep[for a review of SN Ia explosion models
see][]{hillebrandt2000a}. From non-degenerate companions they accrete
matter until 
they approach the Chandrasekhar mass. After some thousand years of
convective carbon burning thermonuclear reactions in the
center of the star form a flame, i.e.~a small spatial region undergoes
thermonuclear runaway. This flame can travel outward in two modes. If
it is mediated by shock waves, the propagation proceeds at
supersonic velocities leading to so-called detonations, while a
mediation due to heat conduction of the degenerate electrons leads to
subsonic deflagrations. A prompt detonation can be excluded as a
model for SNe Ia since it fails to produce the observed composition of
the ashes \citep{arnett1969a}. It burns the entire star at high
densities producing mainly iron group elements.

Consequently a mechanism has to be developed wherein the star can
expand prior to incineration. This is provided if the flame
propagation starts out as a deflagration. This
flame propagation mode, however, is much to slow to explain SN Ia
explosions. Realizing that the astrophysical scenario is characterized
by Reynolds numbers as high as $10^{14}$, one expects the development
of strong turbulent features in the flow. These will interact with the
flame wrinkling and stretching its surface. Since the fuel consumption
rate of deflagration flames is determined by the flame
surface area, turbulence will significantly accelerate the flame
propagation. Three-dimensional SN Ia models
\citep[e.g.][]{hillebrandt2000b} taking this effect into
account have been shown to lead to robust explosions and reproduce
many features of SNe Ia (e.g.~\citealt{reinecke2002d,
gamezo2003a}). There have been speculations on the
possibility of a later transition from the deflagration to a
detonation, but a viable mechanism to trigger this transition in SNe
Ia explosions could not be identified yet \citep{niemeyer1999a}.
In any case, the deflagration stage requires thorough investigation
especially in the context of asymmetries of the explosion
process for it is intrinsically anisotropic.
Due to buoyancy (Rayleigh-Taylor)
instabilities the large scale features of the flame show rising
``bubbles'' of hot and light ashes and in between downdrafts of cold
and dense fuel. It is still an open question, how strong the resulting
asymmetries may become. This issue will be addressed in the present
study.

It is clear that in order to explore asymmetries of the explosion
process it is mandatory to
simulate the full star. Most of the three-dimensional simulations so
far have been restricted to
one spatial octant assuming mirror symmetry to the other
octants. Obviously, this approach imposes an artificial symmetry
constraint on the flame
evolution and our goal in this work is to study the effects that result
from removing this symmetry. There has been only one previous
simulation of the deflagration phase in the full star
\citep{calder2004a}, but it failed to explode. The reason for this
is presumably that it was started with very artificial conditions for
the initial flame (a perfect sphere). 
Nevertheless, the merit of this model is that it
points to the importance of the initial flame configuration in the
context of asymmetries of the explosion process.

The flame ignition is difficult to address in
numerical simulations and afflicted with large uncertainties
\citep{woosley2004a}. We will explore the impact of initial flame
configurations on the explosion models in a parametrized way. As a
first step, in this study we investigate anisotropies in the flame
evolution resulting from simulations covering the full star instead of
only one octant and from moderately asymmetric initial flame
configurations. This allows comparison with previous simulations.
Detailed studies of multi-spot ignition scenarios and
strongly dipolar ignited models are subject to forthcoming
publications.

\section{Implementation}

\subsection{Flame Model}

The stratification of the hot and light ashes below cold and dense fuel
leads to a buoyancy instability known as the Rayleigh-Taylor
instability. In its non-linear stage, bubbles of burning material rise
into the fuel with a velocity \citep{davies1950a}
\begin{equation}
\label{davies_eq}
v \propto \sqrt{g \, L \,\mathit{At}},
\end{equation}
where $g$ and $L$ denote the gravitational acceleration and the length
scale of the bubble, respectively. The Atwood number $\mathit{At}$
characterizes the density contrast between the interior and the
exterior of the bubbles. At the interfaces of the burning bubbles
shear flows give rise to a secondary 
instability, the Kelvin-Helmholtz instability. It generates
turbulent velocity  fluctuations at this (integral) length scale
$L$, which decay to smaller scales in a turbulent
cascade over a so-called inertial range and are dissipated at the
Kolmogorov length. For the velocity fluctuations $v$ at some scale $l$ in the
inertial range Kolmogorov-scaling yields
\begin{equation}
\label{kolmogorov_eq}
v(l) = v(L) \left(\frac{l}{L} \right)^{1/3}.
\end{equation}
The resulting turbulent motions interact with the flame defining the
deflagrations in thermonuclear supernova explosions as a problem of
turbulent combustion.

The correct modeling of turbulent combustion seems to be the crucial
issue in deflagration models of SNe Ia. These can
greatly benefit from the substantial progress that has been made in
turbulent combustion research over the past decades. A fundamental
notion is that turbulent combustion proceeds in different regimes
(see \citet{damkoehler1940a} and \citet{borghi1985a,peters1986a,peters2000a} for
more recent work). The so-called \emph{flamelet regime} of turbulent
combustion applies to most parts of the deflagration in thermonuclear
supernovae.

In this regime, turbulent eddies wrinkle and stretch the
flame but do not penetrate its internal structure. The internal flame
structure (which extends to less than a centimeter) as well as the
smallest scales of flame interaction with turbulence are tiny compared with the
dimensions of a WD star ($\sim$$10^8 \, \mathrm{cm}$). It is therefore
well justified to treat the resulting ``turbulent flame brush'' as a
discontinuity (the so-called \emph{flame front}) separating the fuel
from the ashes, as long as the small-scale effects are appropriately
modeled.

Two approaches are taken to ensure that. First, known effects have to
be taken into
account. This applies mainly to the interaction of the flame with
turbulence on unresolved scales. Wrinkling will increase its surface
and therefore enhance the burning rate. The shortcoming in resolving
the flame surface can be compensated by assigning an increased
\emph{turbulent burning velocity} to the flame front reproducing the
correct fuel consumption rate. It turns out that in the flamelet
regime the turbulent burning velocity
is proportional to the turbulent velocity fluctuations and decouples
completely from the laminar burning speed determined by the
microphysics. Second, one
may suspect that on unresolved 
scales new physical effects occur (like ``active turbulent
combustion'', \citealt{niemeyer1997b}) that affect the flame
propagation on large scales. This has to be studied in separate
small-scale models focusing on a narrow window in scale
space. Simulations of the cellular burning regime \citep{roepke2003a,
  roepke2004a, roepke2004b,bell2004a} led to the conclusion that 
the resulting increased burning velocity can be ``renormalized''
with little effect on the large scale flame propagation in most stages
of the SN Ia explosion.
At least from this regime drastic effects are unlikely. 

Due to the expansion of the WD, the density at which burning takes
place steadily decreases. This is accompanied by a gradual broadening
of flame width. At some point turbulent eddies can affect the
flame's internal structure and burning enters the so-called
distributed regime. Since this is expected to happen at very
late stages, it may not contribute much to the production of iron
group nuclei. Nevertheless, it could significantly increase the
abundances of intermediate mass elements and hence the total energy release
with consequences for modeling of lightcurves and spectra.
This regime is not yet accessible in our code but current efforts aim
at including it in future simulations. In this respect the results
presented here are preliminary.

To summarize the turbulent flame propagation in thermonuclear
supernova explosions we note that large-scale features are driven by
buoyancy according to Eq.~(\ref{davies_eq}). This sets the integral scale of
a turbulent cascade to smaller scales where the velocity fluctuations
in the simplifying picture of Kolmogorov scaling follow
Eq.~(\ref{kolmogorov_eq}). These velocity fluctuations determine the
turbulent flame velocity which thus falls off slower toward
smaller scales than the square root law of Eq.~(\ref{davies_eq}).

\subsection{Numerical methods}

\begin{figure*}[t]
\centerline{
\includegraphics[width = \linewidth]
  {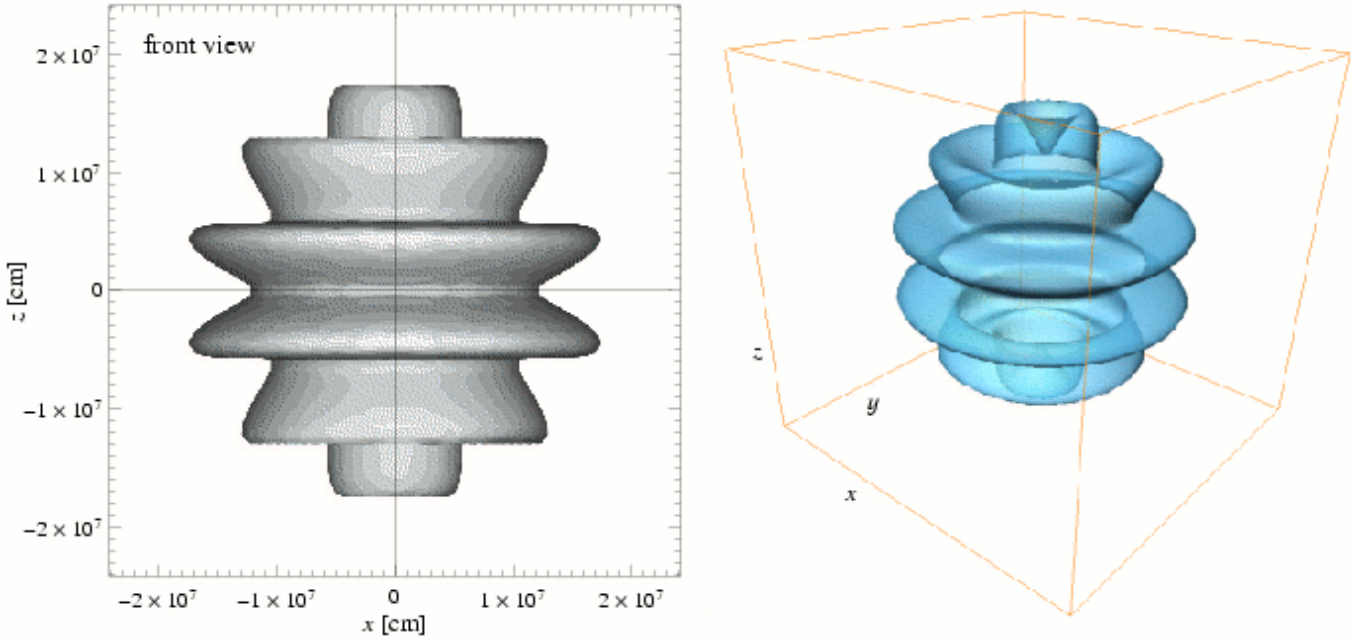}
}
\caption{Initial flame shape of model \emph{c3\_4$\pi$}. \label{iniflame_c3_fig}}
\end{figure*}

\begin{figure*}[t]
\centerline{
\includegraphics[width = \linewidth]
  {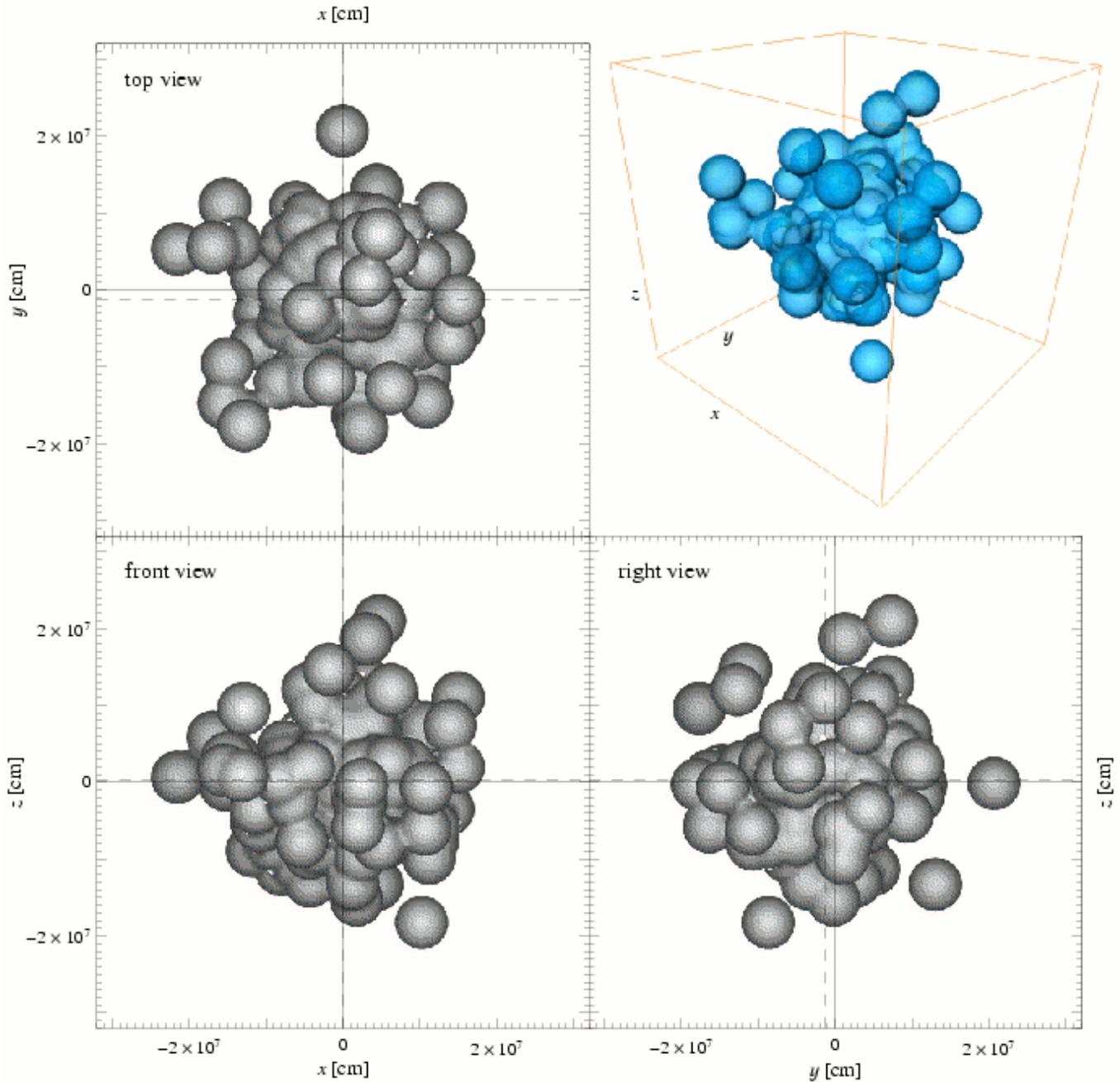}
}
\caption{Initial flame shape of model \emph{f1}. In the projections,
  the solid lines indicate the center of the WD, while the dashed
  lines indicate the center of the flame configuration (the
  $x$-coordinates coincide at the chosen scale of the plot). \label{iniflame_foam_fig}}
\end{figure*}

The implementation of our astrophysical model in a numerical scheme
uses the techniques described by
\citet{reinecke1999a,reinecke2002b}. Only a brief outline will be
given here. 

The hydrodynamics is modeled by
the reactive Euler equations. Their numerical treatment is based on
the \textsc{Prometheus} implementation \citep{fryxell1989a} of the
Piecewise Parabolic Method \citep{colella1984a} and augmented by an
equation of state for WD matter. The description of the nuclear
reactions follows the simplified scheme proposed by
\citet{reinecke2002b} and comprises five species: $\alpha$-particles,
$^{12}$C, $^{16}$O, ``Mg'' as a representative of intermediate mass
elements, and ``Ni'' representing the iron group.

The propagation of the turbulent deflagration flame is modeled
applying the \emph{level set technique} \citep{osher1988a}. Its basic
idea is to associate moving fronts with the zero level set of a signed
distance function $G$, which is then propagated in an appropriate
way. A modified version of this technique has been successfully used to
model terrestrial deflagration flames \citet{smiljanovski1997a}. We
apply the so-called passive implementation of the method \citep[for
details see][]{reinecke1999a}. The propagation of the flame front via
level sets requires knowledge of the flame propagation
velocity, which is composed of two parts: the advection of the flame in
the flow and its self-propagation due to burning. The former is
handled by the hydrodynamics scheme while for the latter the $G$-field
is updated according to the turbulent burning velocity. To
determine this quantity we apply a sub-grid scale turbulence model
first adopted to SN Ia simulations by \citet{niemeyer1995b}.

Our approach is similar to Large Eddy Simulations (LES) known from
computational fluid dynamics.
The efficiency of the numerical scheme is a result of the combination
of level-set techniques with a sub-grid scale model. It enables
numerically converged simulations at rather crude resolutions
\citep{reinecke2002b,reinecke2002c,roepke2004d}.
This makes parameter studies based on three-dimensional simulations
possible \citep[e.g.][]{roepke2004c}.

Besides the wide range of spatial scales, the significant expansion of
the WD during the explosion complicates numerical simulations of
thermonuclear supernovae.
To capture this expansion, the simulations by
\citet{reinecke2002b,reinecke2002d,reinecke2002c} 
were carried out on a nonuniform Cartesian computational grid with
fine resolution in the inner parts and an exponentially expanding grid
spacing in the outer parts of the domain. This, however, causes very low
resolution of the flame front in late stages, especially in directions
diagonal to the axes. To overcome this problem, \citet{roepke2004d}
developed a scheme to simulate thermonuclear supernovae on an
non-static uniform Cartesian computational grid which tracks the
expansion of the WD. Besides a number of numerical advantages
it offers a possibility to follow the evolution of
the explosion to homologous expansion with only moderate additional
computational expenses. Furthermore, at a given time, it provides a
constant resolution throughout the domain. This is important in
full-star simulations without artificial
symmetries. On a non-uniform grid main features of the flame front
could easily propagate in regions with poor spatial resolution. This
is avoided in the \emph{co-expanding uniform grid approach}. For
detailed description and thorough testing of this scheme we refer to
\citet{roepke2004d}. 

\subsection{Simulation setup}

The setup used in the simulations presented below is similar to
that described by 
\citet{roepke2004d}, except for the computational domain now covering
the full WD star instead of only one octant. All simulations were
carried out on a co-expanding uniform Cartesian computational grid of
$[512]^3$ cells. The initial resolution was chosen to be $\Delta x =
7.9 \, \mathrm{km}$. 
Reflecting boundaries were applied on all sides
of the domain. 

The simulations started with a cold isothermal WD star of $1.406 \,
M_\odot$ with a central density of $\rho_c = 2.9 \times 10^9 \,
\mathrm{g}\,\mathrm{cm}^{-3}$ and a temperature of $T_0 = 5 \times
10^5 \, \mathrm{K}$. Its initial composition was chosen as a 50\%
mixture of carbon and oxygen. In this configuration we placed the
initial flame by instantaneously burning the material behind the
front.

\subsection{Initial flame configuration}

A crucial parameter of thermonuclear supernova simulations is the
prescription of the flame ignition. This problem is, however,
difficult to tackle in explosion simulations. The formation of the
initial flame front(s) is determined by about a thousand years of
convective burning prior to thermonuclear runaway. There have been
only few attempts to investigate the ignition conditions
\citep{garcia1995a, woosley2004a} and these conclude that flame
formation is likely to proceed in a number of burning ``bubbles'' that
may rise to distances up to $200 \, \mathrm{km}$ from the center
before the runaway.

Initial flame configurations have to be represented on the rather
coarse computational grids used in
multidimensional thermonuclear supernova simulations. Several
possibilities for this have been suggested in literature. In the simplest
approach one starts out with a flame front in form of a single sphere
\citep[e.g.][]{gamezo2003a}. In this case, however, the high initial
symmetry is broken by numerical noise and it takes a substantial
amount of computer time until the first Rayleigh-Taylor features
develop. This is probably not a very realistic scenario, since the
multitude of burning bubbles predicted by \citet{woosley2004a}
provides seeds for the development of nonlinear buoyancy instabilities
from the very beginning of the flame propagation. Therefore one may start out
with a sphere on which perturbations are imposed, such as the
\emph{c3} model of \citet{reinecke2002b}.
A better approximation is provided by
multi-spot ignition scenarios \citep{reinecke2002d}, where some
burning bubbles are initialized which, due to the lack of resolution,
represent a collection of real flame spots. Here, the
problem is, however, that a rather high resolution is required to
accommodate a decent number of burning bubbles (note that the high
resolution is not necessarily required for numerical
convergence). The models in this paper comprise 
$[256]^3$ computational cells per octant rendering reasonable
multi-spot ignition scenarios impossible. Therefore we
chose to represent the initial flames in one of our simulations by a
cluster of bubbles. These
overlap in the center while some of them float detached from the bulk. The
result is a clumpy or foam-like initial flame structure which we think
provides a reasonable approach given the restricted computational
resources.

Since the initial flame configurations turn out to be a crucial
parameter for the simulations, we will now describe our models
in detail.

As a standard test case \citet{reinecke2002b} established the \emph{c3}
model which is very simple and allowed the comparison of two- and
three-dimensional simulations
\citep[see][]{reinecke2002b}. Here the initial flame is represented by
a sphere which is perturbed by a superposition with toroidal
rings. This configuration is visualized in Fig.~\ref{iniflame_c3_fig},
where the isosurface corresponds to
$G = 0$ representing the flame front. We will denote the full-star
version presented here \emph{c3\_4$\pi$}. Obviously, the center of the
initial flame coincides with the center of the WD in this model.

\begin{figure}[t]
\centerline{
\includegraphics[width = 0.9 \linewidth]
  {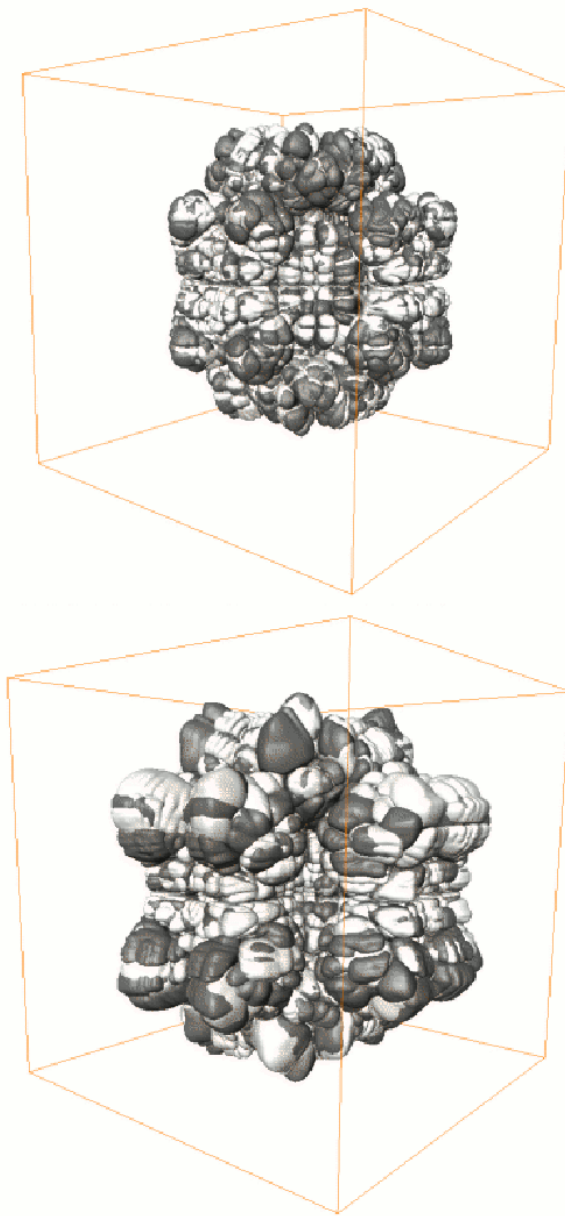}
}
\caption{Comparison of the \emph{c3\_4$\pi$} full star simulation
  (light gray) with the \emph{c3\_oct}
  single-octant simulation mirrored to the other octants (dark
  gray) at $t = 1.5 \, \mathrm{s}$ with an edge length of the cube of $2.16
  \times 10^9 \, \mathrm{cm}$ (top) and at $t = 10.0 \, \mathrm{s}$
  with and
  edge length of the cube of $2.40 
  \times 10^{10} \, \mathrm{cm}$ (bottom) \label{c3_compare_fig}}
\end{figure}

\begin{figure*}[t]
\centerline{
\includegraphics[width = \linewidth]
  {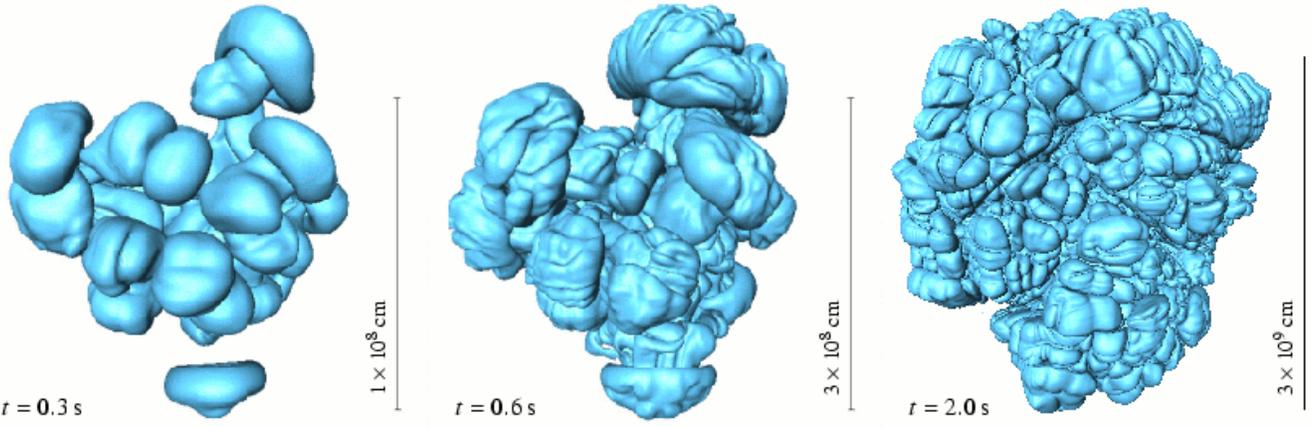}
}
\caption{Flame evolution in model \emph{f1\_512}. \label{foam_evo_fig}}
\end{figure*}

\begin{figure*}[t]
\centerline{
\includegraphics[width = \linewidth]
  {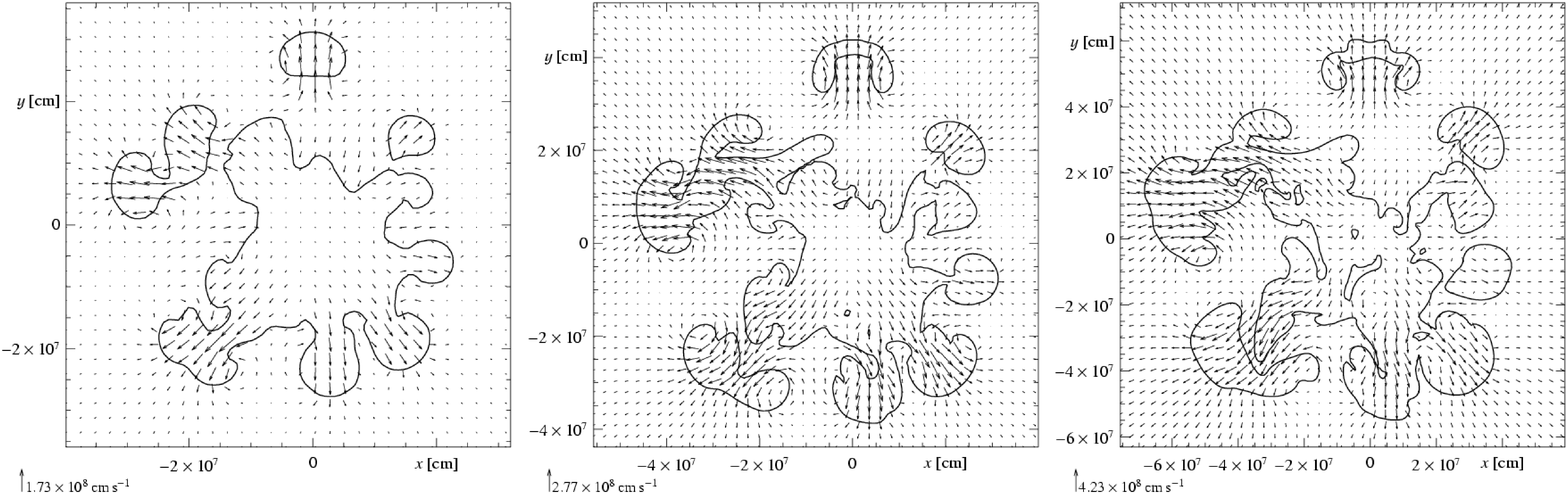}
}
\caption{Velocity fields in two-dimensional slices of the \emph{f1}
  model at $t = 0.1 \,\mathrm{s}$ (left), $t = 0.2 \,\mathrm{s}$
  (center), and $t = 0.3 \,\mathrm{s}$ (right) at $z = -4.0 \times
  10^5 \, \mathrm{cm}$,  $z = -4.0 \times 10^5 \, \mathrm{cm}$, and
  $z = -4.4 \times 10^5 \, \mathrm{cm}$,
  respectively. \label{velslice_fig}}
\end{figure*}

\begin{figure*}[t]
\centerline{
\includegraphics[width = \linewidth]
  {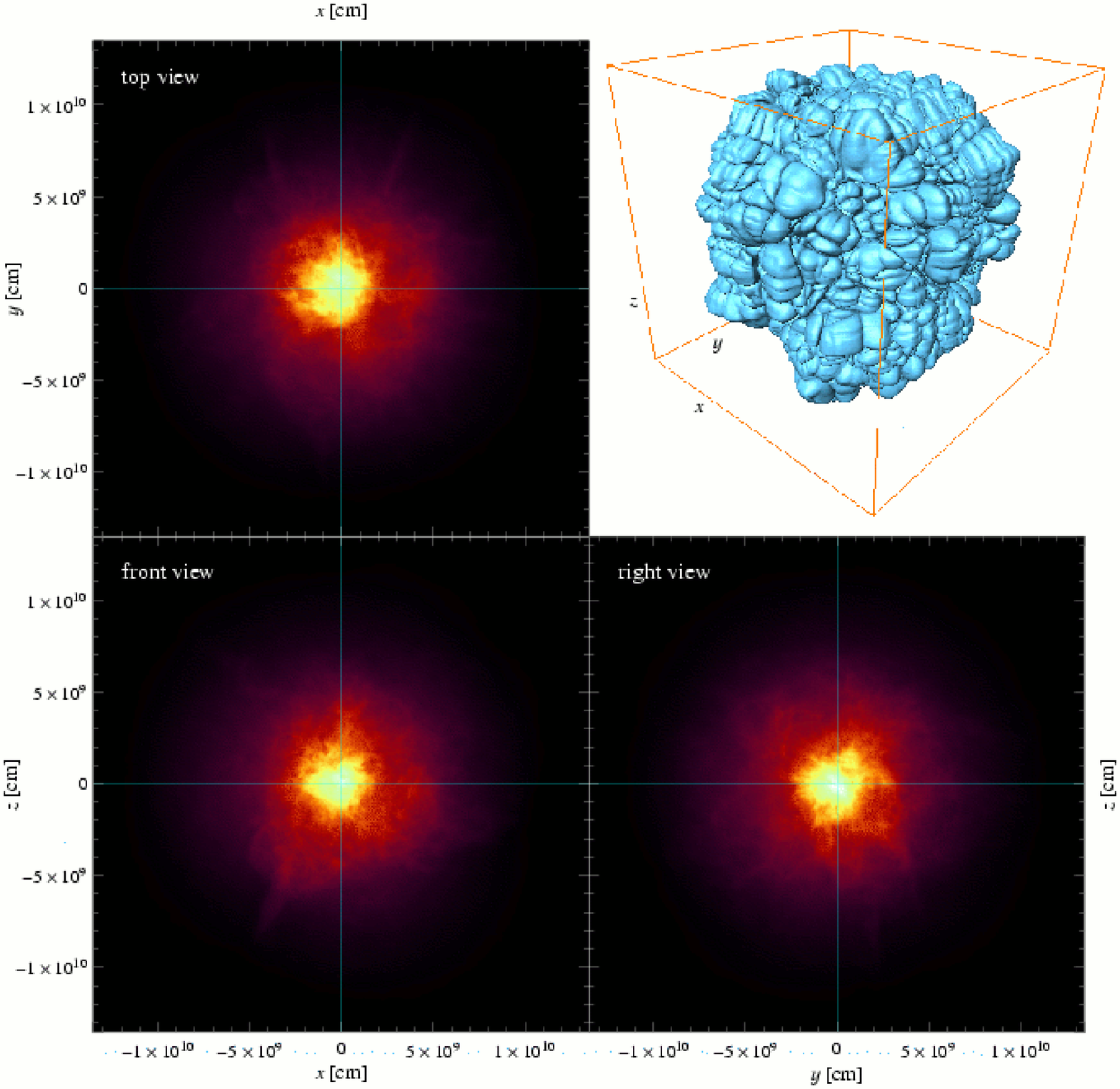}
}
\caption{Model \emph{f1} at $t = 10 \, \mathrm{s}$. In the projections
  the column mass density is linearly color-coded.\label{proj_foam_fig}}
\end{figure*}

A second initial flame configuration was obtained by the superposition
of a number of bubbles around the center of the WD star. The radius of
the bubbles was set to $3.5 \, \mathrm{km}$ and
their locations were chosen in a random selection process under the
constraint of a Gaussian distribution around the center with a
dispersion of $\sigma = 100 \, \mathrm{km}$. The resulting configuration is depicted
in Fig.~\ref{iniflame_foam_fig}. We will refer to the corresponding model
as \emph{f1} in the following. Note that the flame distribution
resulting from the random selection of locations is not symmetric and
that the outermost flame spots are located at a distance of
$\sim$$200 \, \mathrm{km}$ from the center in agreement with
\citet{woosley2004a}. Resulting from the random choice of a limited
number of ``bubbles'', the center of the initial flame is displaced
by $\mbf{r} = (0.3, -13.0,
3.3)\,\mathrm{km}$ from the center of the WD, i.e.\ it is located at a
distance of $13.4 \, \mathrm{km}$ off-center. 

\section{Morphology of the flame evolution}

Analyzing the results of our simulations, we will first describe the
flame evolution in our models. Global characteristics of the
explosions will be discussed in Sect.~\ref{char_sec}.

\subsection{Effects from removing the symmetry constraints}

\citet{roepke2004d} applied a \emph{c3} flame ignition condition for a
demonstration of the co-expanding grid approach in three
dimensions. There only one octant was simulated assuming mirror
symmetry to the other octants. We will now compare this simulation
(denoted \emph{c3\_oct}) with our \emph{c3\_4$\pi$} full-star
simulation. Although the \emph{c3} initial flame shape is not a
realistic scenario, this model offers a way to assess possible
differences in the full star implementation. 

A comparison of the flame shapes at $t = 1.5 \, \mathrm{s}$ and at  $t
= 10.0 \, \mathrm{s}$ is given in
Fig.~\ref{c3_compare_fig}. The isosurfaces again represent $G = 0$,
which is associated to the flame front at early times, while at later
times when the burning has ceased it still provides a good
approximation to the interface between fuel and ashes
\citep[cf.][]{roepke2004d}. Fig.~\ref{c3_compare_fig} shows that
the evolutions of the flame structures of the
\emph{c3\_oct} simulation (here mirrored to the other octants) and the
full-star model proceed very similar. The $G=0$ isosurface in
the full-star simulation is in close agreement with the octant
symmetry of the \emph{c3\_oct} model, even after an evolution time of
$10 \, \mathrm{s}$. 
Obviously, the axial symmetry of the initial flame front is broken in
the same way
in the two implementations. It turns out, that this
happens preferentially at locations where the flame front is aligned
diagonal to the
computational cells of the Cartesian grid indicating that the
symmetry breaking is
induced by discretization effects. 

It seems appropriate to point out that the deflagration SN Ia model
does not develop new large-scale features when the computational
domain is extended to the full star. Unlike convective
burning phenomena (e.g.\ pre-ignition evolution of a WD toward a SN
Ia, see \citealt{woosley2004a}, or core-collapse supernovae, see
\citealt{scheck2004a}) our models form no global
dipole-asymmetries. The reason for this is that the rapid propagation
of the burning front does not allow several eddy turnovers which seem
to be crucial for the development of such configurations. We also find no
indication for the development of larger plume structures in the
\emph{c3\_4$\pi$} simulation than in the \emph{c3\_oct} model. 

Thus, as long as the initial conditions (especially the initial flame
configurations) do not violate the mirror symmetry, it is safe to
restrict the simulations to a single spatial octant.
Also for our full-star models the result is
encouraging since it proves that they are able
to preserve initial symmetries, without explicitly imposing those by
symmetries of the domain. This verifies our full-star
implementation. Hence asymmetries as described below are of physical
nature and not a numerical artifact.

\subsection{Effects of asymmetric initial flame configurations}

We will now address the question how the SN Ia explosion proceeds if
the initial flame configuration does not possess any
symmetry. This is the case for our \emph{f1} model. Starting from the
initial flame configuration shown in Fig.~\ref{iniflame_foam_fig}, the
evolution of the flame front in the explosion process is illustrated
by snapshots of the $G=0$ isosurface at $t = 0.3 \, \mathrm{s}$, $t =
0.6 \, \mathrm{s}$, and $t = 2.0 \, \mathrm{s}$ in
Fig.~\ref{foam_evo_fig}. 

The development of
the flame shape from ignition to $t = 0.3 \, \mathrm{s}$ is
characterized by the well-known ``mushroom-like'' structures resulting
from buoyancy. This is especially well visible for the bubbles that
were detached from the bulk of the initial flame. But also the
perturbed parts of the contiguous flame closer to the center develop
nonlinear Rayleigh-Taylor like features. This flame evolution
generates strong vortices in the resolved flow field, illustrated
by the temporal evolution of the velocities in a two-dimensional
slice in Fig.~\ref{velslice_fig}. The order of magnitude of the
resulting velocities is in 
agreement with Eq.~(\ref{davies_eq}), given the length scale of the
buoyant structures of $\sim$$100\,\mathrm{km}$, a gravitational
acceleration of about $10^{10}\,\mathrm{cm}\,\mathrm{s}^{-2}$, and an
Atwood number of the order of $0.1$. Due to
the multitude of ascending structures, the flow pattern
is strongly irregular.

During the following flame evolution inner structures of smaller
scales catch up with the outer ``mushrooms'' and the separated structures
merge forming a more closed configuration (see snapshot at $t = 0.6 \,
\mathrm{s}$ of Fig.~\ref{foam_evo_fig}). This is a result of the
large-scale flame advection in the turbulent flow and the expansion of
the ashes.

Up to this stage the flame was strongly anisotropic.
However, in the later evolution a preferentially lateral growth of
bubbles filled with ashes smooths out parts of the anisotropies.
As this effect continues to act even after the burning has ceased
in our model, it is of hydrodynamic origin rather than caused by flame
propagation.
The flame develops a more spherical shape and only a slight anisotropy is
retained. This is apparent
from the snapshot at $t = 2.0 \, \mathrm{s}$ in
Fig.~\ref{foam_evo_fig}. At the stage depicted there self-propagation
of the flame due to burning has terminated.

The subsequent evolution is characterized by the approach to
homologous expansion which is reached to a good approximation at $t
\sim 10\, \mathrm{s}$ \citep{roepke2004d}. A snapshot corresponding to
$t = 10\, \mathrm{s}$ is shown in Fig.~\ref{proj_foam_fig}. The
differences to the  snapshot at $t =
2.0 \, \mathrm{s}$ in Fig.~\ref{foam_evo_fig} are small, but the
smoothing effect has rounded the structure somewhat further. 
This confirms the assumption that it is caused by the hydrodynamics.

It is interesting to note that the \emph{f1} model, though it possesses
a slight large-scale asymmetry, finally develops a smoother surface of
ashes than the \emph{c3\_4$\pi$} model (compare the snapshots at $t =
10 \, \mathrm{s}$ in Figs.~\ref{c3_compare_fig} and
\ref{proj_foam_fig}). This can be attributed to the fact that the
amplitude of large-scale perturbations (interpreted as deviations from a
spherical shape) in the \emph{f1} initial flame is smaller,
although it is more perturbed on small scales. 

An important aspect of full-star simulations are the resulting
asymmetries in the density distribution. 
An impression of these is given by the projections of the density to
the coordinate planes in Fig.~\ref{proj_foam_fig}.
The central high-density region is aspherical and also in the outer
regions filaments of increased density are
visible. These result from downdrafts of dense fuel between
bubbles of light ashes during the explosion.

Despite the slight initial misalignment of the centers of the WD and
the flame, we observe no buoyancy-driven ascension of the flame as a whole
toward the surface. For this a large-scale flow around the flame
structure would be necessary. Such a flow-structure can, however, not
develop since our initial flame configuration causes an irregular
vortical velocity field, which dominates the flame advection. We
emphasize here, that this large-scale effect is not caused by the
self-propagation of the flame and should thus be independent of the
particular turbulent flame model.

\begin{figure}[t]
\centerline{
\includegraphics[width = \linewidth]
  {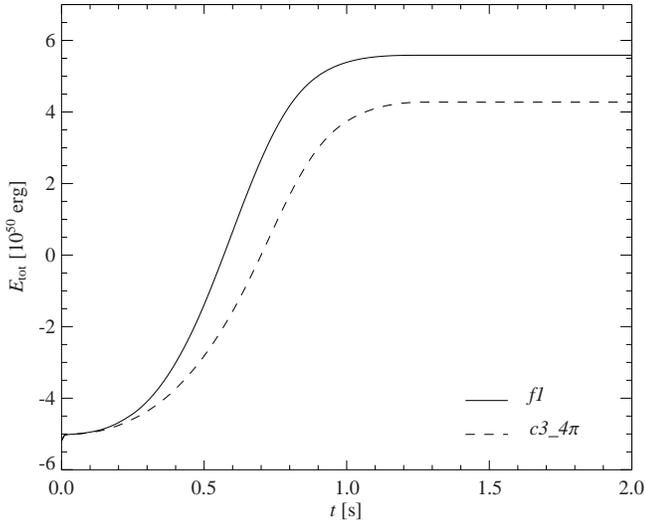}
}
\caption{Temporal evolution of the total energies in our
  models. \label{etot_fig}}
\end{figure}

\begin{figure}[t]
\centerline{
\includegraphics[width = \linewidth]
  {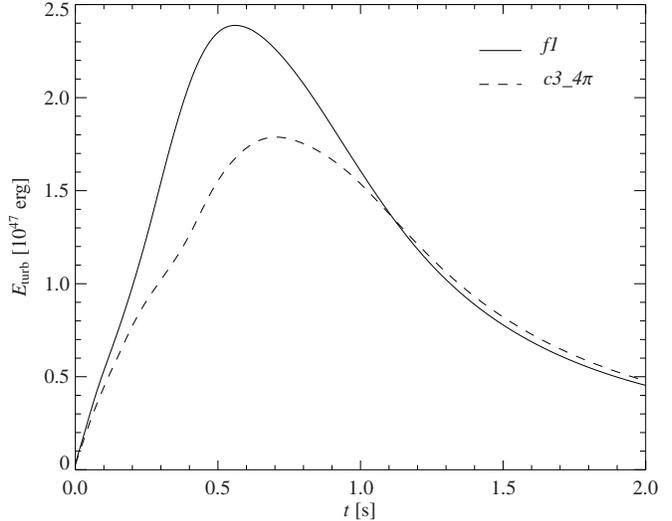}
}
\caption{Temporal evolution of the turbulent energies in our
  models. \label{eturb_fig}}
\end{figure}

\begin{table}
\centering
\caption{Global characteristics of the models.
\label{global_tab}}
\setlength{\extrarowheight}{2pt}
\begin{tabular}{p{0.08 \linewidth}p{0.16 \linewidth}p{0.16 \linewidth}p{0.16 \linewidth}p{0.16 \linewidth}}
\hline\hline
model & $m\,$(``Ni'') $[M_\odot]$ & $m\,$(``Mg'') $[M_\odot]$
& $E_\mathrm{tot}$ $[10^{50} \, \mathrm{erg}]$ & $E_\mathrm{nuc}$ $[10^{50} \, \mathrm{erg}]$ 
\\
\hline
\emph{c3\_4$\pi$} & 0.507 & 0.151 & 4.275 & 9.466 \\
\emph{f1}         & 0.579 & 0.165 & 5.585 & 10.776 \\
\hline
\end{tabular}
\end{table}

\section{Global parameters of the explosion models}
\label{char_sec}

The global characteristics of the models are summarized in
Table~\ref{global_tab}. It is obvious that the \emph{f1} initial
flame configuration leads to a more vigorous explosion than the
\emph{c3\_4$\pi$} model. It produces 14\% more iron group elements and
consequently the release of nuclear energy is 14\% higher. This
amounts to a 31\% increase in the total explosion energy. The excess
in produced intermediate mass nuclei is only 9\% and thus the
relative fraction of these is lower in
the \emph{f1} model. The temporal evolutions of the total energies in
our simulations are plotted in Fig.~\ref{etot_fig}.

Two aspects of the \emph{f1} model contribute to the more energetic
explosion. First,
the initial flame configuration extends to slightly larger radii than
the \emph{c3\_4$\pi$} initial flame (compare
Figs.~\ref{iniflame_c3_fig} and 
\ref{iniflame_foam_fig}). Therefore the outer flame fronts experience
a larger gravitational acceleration and the development of structures
resulting from buoyancy instabilities is accelerated. This leads
to an enhanced generation of turbulence and consequently the value of
the turbulent flame propagation velocity is higher. Hence more
material is burnt at higher densities resulting in iron group
nuclei. 

Second, due to its composition of small bubbles the initial flame
shape of the \emph{f1} model is perturbed with shorter wavelengths.
This provides more seeds for the development of buoyancy
instabilities and consequently the wrinkling of the flame is is
enhanced resulting in an increased fuel consumption.
Contrary to that, in the \emph{c3\_4$\pi$} model it
takes some time until the axial symmetry is broken and ``mushroom-like''
structures form. The \emph{f1} model therefore seems more appropriate
to mimic realistic initial flame structures (perturbed by
convective motions and eventually igniting in multiple spots) on coarse
computational grids. The evolutions of the turbulent energies in both
simulations are plotted in Fig.~\ref{eturb_fig}, showing clearly the
enhanced turbulence production in the \emph{f1} model. Since the turbulent
burning velocity scales with the turbulent velocity fluctuations, the
flame propagates faster in that model.

Synthetic light curves and spectra derived from explosion models are
largely determined by the distribution of the ejecta in velocity
space. However, the interpretation of these data is not
straightforward and requires caution. One of the former main
uncertainties in analyzing the results from three-dimensional
explosion  simulations, namely the short simulation time, could
be overcome in the
models presented here. On the co-expanding computational grid they
were evolved to $10 \, \mathrm{s}$ providing a fair approximation
to homologous expansion \citep{roepke2004d}. However, most of the
currently available numerical codes treat the radiation transport
one-dimensional. One should keep in mind, that
multidimensional structures  (cf.\ Fig.\ref{proj_foam_fig}) could
considerably change the results from three-dimensional synthetic light
curve and spectra calculations depending on the line of sight.

\begin{figure}[t]
\centerline{
\includegraphics[width = \linewidth]
  {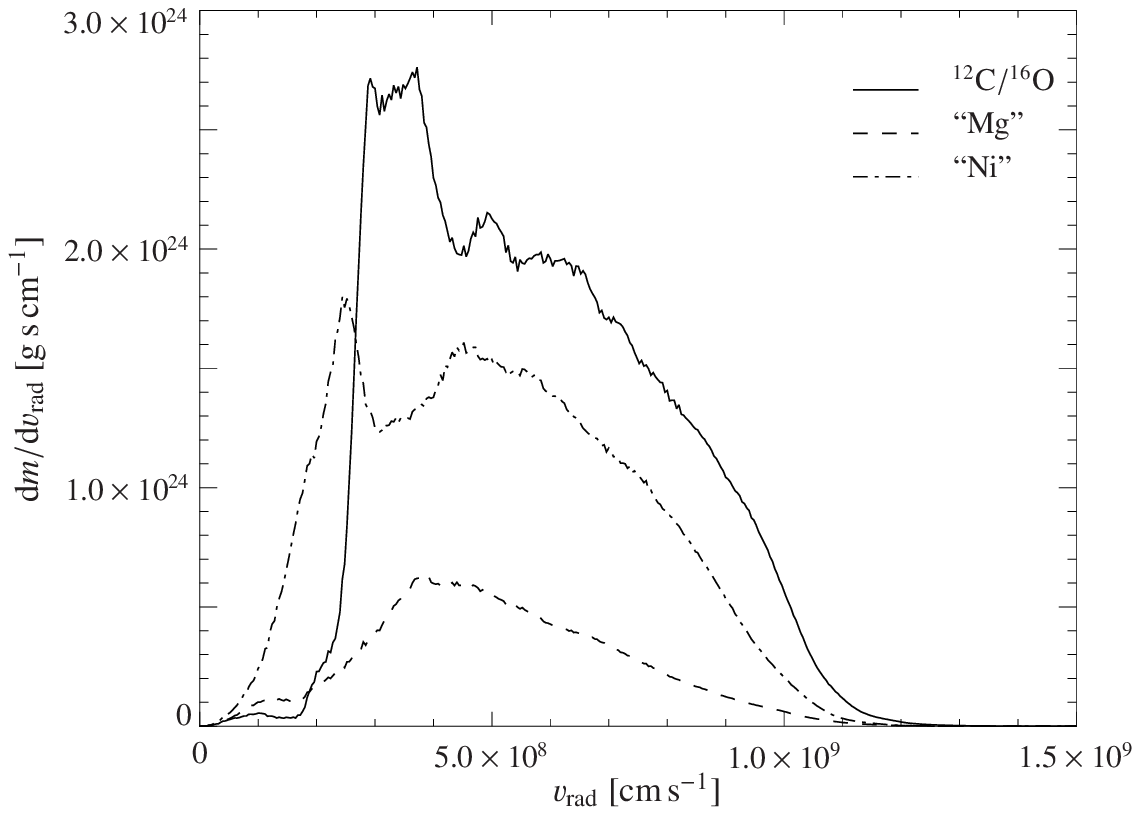}
}
\caption{Angular averaged distribution of the ejecta of model
  \emph{c3\_4$\pi$} in radial velocity space after 10 s. \label{velprofile_fourpi_fig}}
\end{figure}

\begin{figure}[t]
\centerline{
\includegraphics[width = \linewidth]
  {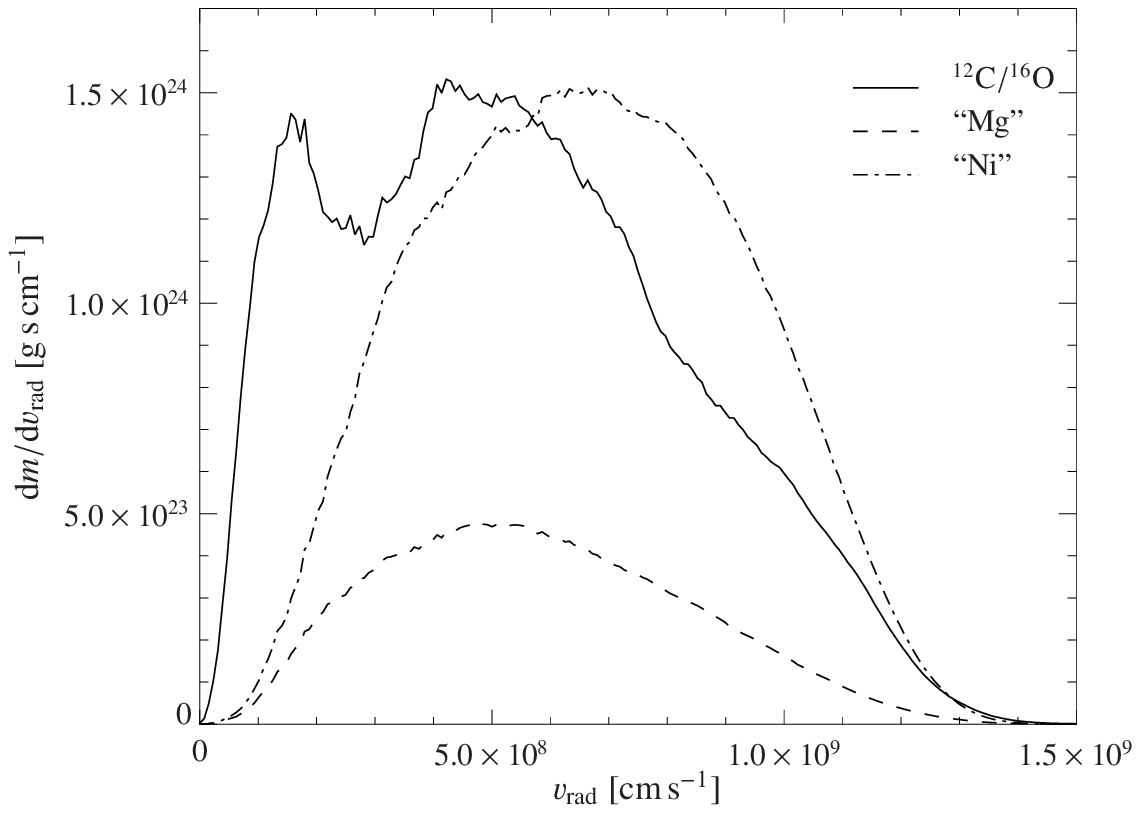}
}
\caption{Angular averaged distribution of the ejecta of model
  \emph{f1} in radial velocity space after 10 s. \label{velprofile_foam_fig}}
\end{figure}

 In Figs.~\ref{velprofile_fourpi_fig} and
\ref{velprofile_foam_fig} we present the angular averaged profiles of
the velocity distribution of the ejecta.
These figures show
clearly that both explosion models are too weak to accelerate the outer
ejecta to the highest observed velocities \citep[see for example the
velocities of the Si \textsc{ii}, S \textsc{ii} and Ca \textsc{ii}
lines given by][]{benetti2004a}. While they reach to
$\sim$$1.2 \times 10^9 \mathrm{cm} \, \mathrm{s}^{-1}$ in the
\emph{c3\_4$\pi$} model, the outermost ejecta of the \emph{f1} model
are found at velocities of $1.5 \times 10^9 \, \mathrm{cm} \,
\mathrm{s}^{-1}$. 

Due to the close similarity of the
\emph{c3\_4$\pi$} simulation to the \emph{c3} model calculated in only one
octant, the distribution of the species in the ejecta shows no visible
differences \citep[compare with the analogous plot for the \emph{c3} model
of][]{roepke2004d}. However the deviations from the the
profiles given by \citet{reinecke2002d} are significant, especially at higher
velocities. This is simply due to the fact, that we evolved the model
to $10 \, \mathrm{s}$, while former simulations stopped at $1.5 \,
\mathrm{s}$ where homologous expansion is not yet reached
\citep{roepke2004d}. 
At $t = 1.5 \, \mathrm{s}$ the distribution of the elements is quite
uniform, but this is not retained in the approach to homologous
expansion. The later accumulation at lower velocities can probably be
attributed to the still non-negligible gravitational potential.

A feature of the \emph{c3} model seems to be that little fuel is
left at the center of the star. For this, a possible explanation would
be that with the given initial flame configuration it takes longer to break
the axial symmetry. True three-dimensional features with
``mushroom-like'' bubbles of ashes and downdrafts of fuel form
later. Therefore less unburnt material sinks down to the
center.

This is different in the species distribution of the \emph{f1}
model shown in Fig.~\ref{velprofile_foam_fig}. Here the higher amount
of unprocessed material near the center can be attributed to the
accelerated development of the typical buoyancy-induced flame
structures. The earlier formation of downdrafts transports unburnt
matter more efficiently toward the center. However, this effect may be
partly compensated when the model is started from a true multi-spot
ignition scenario \citep[cf.\ to the \emph{b30} model
of][]{travaglio2004a}. In this case the increased surface of the flame
near the center leads to a more efficient burning. A possible temporal
distribution of the formation of initial flames would also favor an
enhanced burning near the center.

A second difference between our two models is the distribution of iron
group nuclei (denoted as ``Ni''). The accelerated flame evolution and
the stronger explosion mix these elements further out in the \emph{f1}
simulation. A similar, but weaker effect can be observed for the
intermediate mass nuclei (``Mg''). The low production of intermediate
mass elements in both models is in part due to the small computational
grid applied in the simulations. Naturally, subsequent stages of the
burning are less resolved in the co-expanding grid
approach. By means of a resolution study \citet{roepke2004d} found
that starting with $[256]^3$ cells per octant provides sufficient
resolution in the early stages of flame propagation when iron group
nuclei are synthesized. However, the later production of intermediate
mass elements may be underestimated by about 50\% due to the coarse
resolution of the flame. This, in turn, leads to a somewhat lower
total energy of the models. Also the lower relative production of
intermediate mass elements in the \emph{f1} simulation may partially
be caused by a resolution effect. Since this model explodes more
vigorously, the WD expands faster. The computational grid tracking the
expansion will therefore be coarser at the stages of intermediate mass
element production in the \emph{f1} model than in the less energetic
\emph{c3\_4$\pi$} simulation.

\section{Conclusions}

We presented two full-star simulations of deflagration models for
thermonuclear supernova explosions. These were started with different
initial flame configurations. 
Both simulations resulted in robust explosions, though they were
rather on the weak side compared with observations.
The models allowed an analysis of the effects of
removing the artificial symmetries of former simulations.
The \emph{c3\_4$\pi$} model retained the axial symmetry of our
standard \emph{c3} example although it was calculated on a
full-star domain.
We note that the abolishment of the domain-symmetries alone does not
introduce any
additional asymmetries to the model. This result validates previous
single-octant simulations \citep[e.g.][]{reinecke2002b, reinecke2002d,
  reinecke2002c, roepke2004c} since it shows that the symmetry
constraint imposed there did not suppress physical effects.

To introduce asymmetries into our models it is necessary to start with
an asymmetric initial flame configuration. This was realized by the
``foamy'' initial flame surface of the \emph{f1} simulation.
Starting the flame propagation with such an anisotropic configuration leads
to asymmetries in the ejecta. However, the initial asymmetries are
partially smoothed out by the expansion of the ash-filled bubbles,
which in late phases proceeds preferentially in lateral directions. To
what degree the retained asymmetry could contribute to the
polarization of the spectrum remains to be tested by multidimensional
radiative transfer calculations. 

As a consequence of the perturbed initial flame shape an irregular flow
pattern establishes. This has strong impact
on the flame evolution. In the \emph{f1} simulation two effects are
noted. First, small bubbles in the 
initial flame configuration that were detached from the bulk merge
with it shortly after ignition resulting in a more or less
connected flame structure. Second, the model is robust against slight
misalignments of the centers of the flame and the WD. No buoyancy
driven rise of the flame as a whole toward the surface of the star was
observed. This is in contrast to the result of \citet{calder2004a},
who found that the flame quickly ascends from the center although the
initial misalignment was somewhat smaller than in our \emph{f1}
model. The reason thereof seems to be that \citet{calder2004a}
started with flame in shape of a perfect sphere. This configuration
possesses no seeds for the development of Rayleigh-Taylor instabilities
which consequently have to be introduced by numerical
effects. Therefore the establishment of a vortical velocity field is
delayed and
a symmetric large-scale flow may develop driving the flame to larger
radii. Differences in the flame models
could also effect the results, but seem unlikely to be the primary
reason for the discrepancy. While \citet{calder2004a} use the flame
model of \citet{gamezo2003a} which is based on the assumption of a
buoyancy dominated scaling according to Eq.~(\ref{davies_eq}), we model the
turbulent flame speed taking into account the interaction of the flame
with turbulence on unresolved scales. From Eq.~(\ref{kolmogorov_eq})
it is clear that this results in a slower decrease of the flame speed
toward smaller scales.

In this first study only one anisotropic initial flame configuration
was tested. Open questions remain regarding the flame evolution in
case of stronger initial asymmetries, especially more pronounced
dipole shapes. How robust are the models to stronger displacements of
the initial flame from the center? What asymmetries can then result
from deflagration models? An extreme case would be a one-sided ignition
which was suggested by \citet{woosley2004a}. Such models will be
addressed in a forthcoming study.


\begin{acknowledgements}
We would like to thank M.~Reinecke for helpful discussions. 
This work was supported in part by the European Research
Training Network ``The Physics of Type Ia Supernova Explosions'' under
contract HPRN-CT-2002-00303. We thank the Institute for Nuclear Theory
at the University of Washington for its hospitality during the summer
program on Supernovae and Gamma Ray Bursts (INT-04-2), where this
paper was completed. Discussions with the participants of the summer
program inspired some of the work presented here.
\end{acknowledgements}


\end{document}